# REENTRANT SUPERCONDUCTIVITY IN SUPERCONDUCTOR/FERROMAGNETIC-ALLOY BILAYERS

V.I. Zdravkov[1,2], J. Kehrle[2,*], G. Obermeier[2], S. Gsell[2], M. Schreck[2], C. Müller[2],
H.-A. Krug von Nidda[2], J. Lindner[2], J. Moosburger-Will[2], E. Nold[3], R. Morari[1],
V.V. Ryazanov[4], A.S. Sidorenko[1,3], S. Horn[2], R. Tidecks[2], and L.R. Tagirov[2,5]

[1]*Institute of Electronic Engineering and Nanotechnologies ASM, MD2028 Kishinev, Moldova*
[2]*Institut für Physik, Universität Augsburg, D-86159 Augsburg, Germany*
[3]*Institute of Nanotechnology, Karlsruhe Institute of Technology, D-76021 Karlsruhe, Germany*
[4]*Institute for Solid State Physics, Russian Academy of Sciences, 132432 Chernogolovka, Russia*
[5]*Solid State Physics Department, Kazan State University, 420008 Kazan, Russia*

We studied the Fulde-Ferrell-Larkin-Ovchinnikov (FFLO) like state established due to the proximity effect in superconducting Nb/Cu$_{41}$Ni$_{59}$ bilayers. Using a special wedge-type deposition technique, series of 20-35 samples could be fabricated by magnetron sputtering during one run. The layer thickness of only a few nanometers, the composition of the alloy, and the quality of interfaces were controlled by Rutherford backscattering spectrometry, high resolution transmission electron microscopy, and Auger spectroscopy. The magnetic properties of the ferromagnetic alloy layer were characterized with superconducting quantum interference device (SQUID) magnetometry. These studies yield precise information about the thickness, and demonstrate the homogeneity of the alloy composition and magnetic properties along the sample series. The dependencies of the critical temperature on the Nb and Cu$_{41}$Ni$_{59}$ layer thickness, $T_c(d_S)$ and $T_c(d_F)$, were investigated for constant thickness $d_F$ of the magnetic alloy layer and $d_S$ of the superconducting layer, respectively. All types of non-monotonic behaviors of $T_c$ versus $d_F$ predicted by the theory could be realized experimentally: from reentrant superconducting behavior with a broad extinction region to a slight suppression of superconductivity with a shallow minimum. Even a double extinction of superconductivity was observed, giving evidence for the multiple reentrant behavior predicted by theory. All critical temperature curves were fitted with suitable sets of parameters. Then, $T_c(d_F)$ diagrams of a hypothetical F/S/F spin-switch core structure were calculated using these parameters. Finally, superconducting spin-switch fabrication issues are discussed in detail in view of the achieved results.

___

*Corresponding author. Tel.: +49 821 598 3407; Fax +49 821 598 3411.

*E-mail address:* jan.kehrle@physik.uni-augsburg.de (J. Kehrle)

# I. INTRODUCTION

Superconductor/ferromagnet (S/F) proximity systems show several unusual physical effects originating from the competition of the two antagonistic long-range orderings [1-4]. Usually, ferromagnetism is expected to suppress singlet superconductivity, as the presence of an exchange-field-induced splitting of the conduction band breaks the time-reversal symmetry of a Cooper pair. Fulde-Ferrell and Larkin-Ovchinnikov (FFLO) showed that, nevertheless, superconductivity may exist in the presence of a magnetic background [5,6], for an extremely narrow range of parameters [7]. However, a FFLO-like state can be realized in S/F layered structures. Using weak ferromagnets, like ferromagnetic alloys [1,2], or partial isolation of the S and F layers in tunnel heterostructures [3] weakens the drastic suppression of superconductivity in the vicinity of a S/F interface.

The finite momentum, which the Cooper pair acquires in the exchange field of the ferromagnet, makes the pairing wave function oscillating. The resulting phase change across the ferromagnetic layer is responsible for the $\pi$-junction effects [8-10]. The interference of the incident and reflected wave functions determines the oscillatory phenomena of the critical temperature $T_c$ versus the F layer thickness $d_F$ in S/F bilayers and multilayers [11-14].

Two ferromagnetic layers offer a further control of the superconducting state in a S layer sandwiched in between, if one allows for the rotation of the magnetization of one of the layers with respect to the other [15-19]. For a thin S layer, with a thickness $d_S$ comparable to the superconducting coherence length $\xi_S$, superconductivity can be switched on and off by rotating the magnetization of one of the magnetic layers in a F/S/F trilayer [17].

A necessary condition to optimize the switching effect is the reentrant behavior of superconductivity versus $d_F$ [20-23]. Recently, we reported on the first convincing (and up to now unique) observation of a pronounced reentrant superconductivity phenomenon in S/F layered systems, using Nb/Cu$_{41}$Ni$_{59}$ bilayers [24]. The main goal of this paper is to present new experimental data, which demonstrate all types of non-monotonic behaviors of the superconducting critical temperature as a function of the ferromagnetic alloy layer thickness predicted by the theory: from very expressed reentrant superconductivity with a second appearing of an interference induced extinguished superconducting state, giving evidence for the predicted multiple reentrant state, over a deep minimum of $T_c$ to the slight suppression of superconductivity with a shallow minimum commonly observed by other authors [25-34]. Moreover, with the material parameters obtained from the fitting of the theory to the experiments on S/F bilayers we give a forecast for the sample design and the behavior of a hypothetical F/S/F spin switch made from Nb/Cu$_{41}$Ni$_{59}$ alloy. The necessary theoretical background for treatment of the experimental data is given in the Appendix.

## II. SAMPLE PREPARATION AND CHARACTERIZATION

### A. Thin film deposition and sample preparation

The S/F samples were prepared by magnetron sputtering on commercial (111) silicon substrates at room temperature. The base pressure in the "Leybold Z400" vacuum system was about $2\times10^{-6}$ mbar. Pure argon (99.999%, "Messer Griesheim") at a pressure of $8\times10^{-3}$ mbar was used as sputter gas. Three targets, Si, Nb and Cu$_{40}$Ni$_{60}$ (75 mm in diameter), were pre-sputtered for 10-15 minutes to remove contaminations. Since Nb acts as a getter material, moreover the residual gas pressure in the chamber reduces. Then, a silicon buffer layer was deposited using a RF magnetron. This was to generate a clean interface for the subsequently deposited niobium layer. To obtain flat, high-quality Nb layers with thickness in the range of 5-15 nm, we rotated the target around the symmetry axis of the vacuum chamber during deposition by DC magnetron sputtering [24]. A dc-motor moved the full-power operating magnetron along the silicone substrate of $80\times7$ mm$^2$ size. Thus, the surface was homogeneously sprayed with the material. The average growth rate of the Nb film was about 1.3 nm/sec. The deposition rate for a fixed, non-moving target would be about 4-5 nm/sec.

As in our previous works [24,35], we next deposited a wedge-shaped ferromagnetic layer utilizing the intrinsic spatial gradient of the deposition rate. The Cu$_{40}$Ni$_{60}$ target was RF sputtered with a rate of 3-4 nm/sec, resulting in practically the same composition of the alloy in the film. To prevent a degradation of the resulting Nb/Cu$_{41}$Ni$_{59}$ bilayers at atmospheric conditions, they were coated by a silicon cap of about 5-10 nm thickness (see inset in Fig. 1 for a sketch of the resulting specimen).

Samples of equal width (about 2.5 mm) were cut perpendicular to the wedge to obtain a batch of S/F bilayer strips, with varying Cu$_{41}$Ni$_{59}$ layer thickness $d_F$, for $T_c(d_F)$ measurements. Aluminum wires of 50 μm in diameter were then attached to the strips by ultrasonic bonding for four-probe resistance measurements.

The samples for the $T_c(d_S)$ measurements were prepared by the same procedure, but now with a Cu$_{41}$Ni$_{59}$ film of constant thickness on top of a wedge-shaped Nb layer. In addition, single flat Nb films and single CuNi-wedge shaped layers were prepared in a similar way for materials characterization.

## *B. Thickness and composition characterization*

Rutherford backscattering spectrometry (RBS) was used to determine the thickness of the Nb and $Cu_{1-x}Ni_x$ layers as well as to check their composition $x$. The applicability of this method has been demonstrated in our previous works [24,35]. It allows to determine the thickness of the layers with an accuracy of ±3% for copper-nickel on the thick end of the wedge. For niobium and for copper-nickel on the thin end of the wedge the accuracy is ±5%.

The measurements were carried out using 3.5 MeV $He^{++}$ ions generated by a tandem accelerator. Backscattered ions were detected under an angle of 170° relative to the incident beam by a semiconductor detector. To avoid channeling effects in the Si substrate, the samples had to be tilted by 7° and azimuthally rotated during measurement. The RUMP computer program was used to simulate the spectra [36]. From the resulting elemental areal densities of Nb, Cu and Ni the thickness of niobium and copper-nickel layers was calculated, considering the densities of the respective metals.

The results for the layer thickness and $Cu_{1-x}Ni_x$ alloy composition $x$ as a function of sample number, *i.e.* the position on the substrate of batch S22 are shown in Fig. 1(a). The Ni concentration $x$ in the $Cu_{1-x}Ni_x$ layer is nearly constant, exhibiting a slight increase from 58 at.% to 60 at.% Ni towards the thick side of the wedge. The thickness of the Nb layer is practically constant along the wedge, $d_{Nb}(S22) \approx 7.8$ nm +0.7nm/-0.8nm.

Figure 1(b) shows a cross-section of a $Si(substrate)/Si(buffer)/Nb/Cu_{41}Ni_{59}/Si(cap)$ sample (#18 from batch S22) displayed by transmission electron microscopy (TEM). According to RBS measurements (linear interpolation between sample #16 and #19 at 36.9 mm and 44.6 mm distance from the thick side of the wedge, respectively) it is $d_{Nb}(RBS) \approx 7.2$ nm and $d_{CuNi}(RBS) \approx 14.1$ nm. From the TEM picture one can determine the Nb layer thickness as $d_S(TEM) \approx 7.4$ nm and the CuNi layer thickness as $d_F(TEM) \approx 15.0$ nm, in good agreement with the RBS data for the sample quoted before.

## *C. Scanning Auger electron spectroscopy*

To study the quality of interfaces between the layers we have performed Auger electron spectroscopy (AES) measurements [37] of Si(substrate)/Si(buffer)/Nb/Si(cap) and Si(substrate)/Si(buffer)/Nb/Cu$_{1-x}$Ni$_x$/Si(cap) specimens. A defocused Xe-ion beam erodes a crater into the film with inclination angles of the scarps of only a few degrees or below. An electron beam then scans the shallow crater. The secondary electrons provide a topographic view (in which the films extend over a much larger region than for a perpendicular cut, due to the small inclination angle of the crater), whereas the emitted Auger electrons reveal the lateral distribution of elements. As a result, one obtains the elemental concentration as a function of sample depth. Our AES data of the Nb specimen in Fig. 2a show that the interface of the substrate with the buffer layer has considerable oxygen content because of the oxidized surface of the silicon wafer. The little amount of carbon is due to carbon dioxide absorbed by the oxidized surface before the deposition process has been initiated. We did not find any detectable contaminations of the Si buffer, Nb, and Nb-Si cap interfaces formed during sputtering. The 10 nm thick silicon cap layer provided a long-lasting protection of the Nb layer against deterioration in the ambient atmosphere (the superconducting $T_c$ shift is less that 0.15 K after 1 year of aging at ambient conditions). The AES data for the Nb/Cu$_{1-x}$Ni$_x$ specimen (Fig. 2b) show similar features. There are about 59.3 at.% Ni (in agreement with RBS data of 59 at.%) and 39.0 at.% Cu in the Cu$_{1-x}$Ni$_x$ film and a small amount of carbon (~1.7 at.%) in addition. There is a small concentration of O, C and N impurities at the Nb/Cu$_{1-x}$Ni$_x$ interface whereas the Cu$_{1-x}$Ni$_x$/Si(cap) interface is free of contaminations.

### *D. Magnetic properties of the ferromagnetic alloy layer*

The magnetic properties along the wedge of a ferromagnetic alloy Cu$_{1-x}$Ni$_x$ layer were measured with a Quantum Design MPMS-5 superconducting quantum interference device (SQUID) magnetometer to examine their homogeneity. From the magnetic moment $m$ [emu], measured by the SQUID, the magnetization $M$ [emu/mol] was calculated by taking the sample geometry into account, *i.e.* $M = m$[emu]/[(V$_{film}$/V$_{mol,CuNi}$)mol]. Here, $V_{film}$ is the volume of the Cu$_{41}$Ni$_{59}$ film, $V_{mol,CuNi}$ = 6.8 cm$^3$ the volume of 1 mol of Cu$_{41}$Ni$_{59}$.

Figure 3 shows the temperature dependence of the magnetization measured during cooling down, for samples #21 and #1 and warm up for sample #12, respectively, with an applied external field of $H$ = 1000 Oersted (where 1 Oe = $(10^3/4\pi)$(A/m) = 79.58 A/m in international (SI) units) [38]. Sample #12, measured upon warming, was cooled down in a field of 1000 Oe before. It should, however, not be essential in which field the sample was cooled down before, because for $H = 1000$ Oe we are at the point were all branches of the hysteresis curve join (see Fig. 4a).

The samples were taken from the thin end, the middle, and the thick end of the $Cu_{41}Ni_{59}$ wedge, respectively. The magnetic field was applied in plane of the film, parallel to the long side of the sample cut from the wedge as described in detail in Sec. II A. The temperature dependence of the magnetization exhibits the paramagnetic to ferromagnetic transition at Curie temperatures which are somewhat lower than $T_C=180$ K for Cu with 59 at°% Ni bulk material [39], namely, at $T_C = 100$ K, 110 K and 118 K for samples #21 ($d_{CuNi} = 8$ nm), #12 ($d_{CuNi} = 21$ nm), and #1 ($d_{CuNi} = 48$ nm), respectively. The thickness dependence of $T_C$ is in agreement with $T_C(d_{CuNi})$ measured for $Cu_{40}Ni_{60}$/Cu multilayers by Ruotolo *et al.* (Ref. [40], Fig. 6) for $d_{Cu} = 10$nm. This result proves the high homogeneity of the magnetic properties along the wedge. Note that the offset of the magnetization above $T_C$ due to the huge diamagnetic background from the silicon substrate and due to the ferromagnetic background (if present) was set to zero.

Figure 4a shows the in-plane hysteresis for one of the samples measured at 2 K. It consists on one hand of a linear contribution with negative slope, resulting from the diamagnetic Si substrate, and on the other hand of a ferromagnetic hysteresis loop of the thin film of the $Cu_{41}Ni_{59}$ alloy. Since the diamagnetic signal of the Si substrate has not been subtracted, the magnetic moment *m(H)* as measured by the SQUID is shown in this figure. The hysteresis shown in Fig. 4a actually is a tiny signal on a large diamagnetic background, as seen in Fig. 4b. This diamagnetic signal would cross the origin if there would be no hysteresis curve present. The combined signal becomes a straight line, too, as soon as the magnetic field is high enough to drive the film into the saturated sate. In this case the magnetic moment is given by $m=\chi_{Si}[(V_{Si}/V_{mol,Si})mol]H+[(V_{film}/V_{mol,CuNi})mol]M_S$. Here, $\chi_{Si}$ [emu/(mol×Oe)], $V_{Si}$ and $V_{mol,Si}$ are the magnetic susceptibility, the volume of the Si substrate and the volume of 1 mol Si, respectively, and $M_S$ [emu/mol] is the saturation magnetization of the $Cu_{41}Ni_{59}$ film. Thus, the back extrapolation of the straight line behavior observed for high applied fields to *H*=0 yields $m_S = [(V_{film}/V_{mol,CuNi})mol]M_S$ (Fig. 4b). For sample WCN3#12, thus, we get $m_S = 3.7\times10^{-5}$ emu yielding $M_S = 790$ emu/mol. For sample WCN3 #1 (hysteresis curves not shown in the present work) the same procedure results in $m_S = 1.1\times10^{-4}$ emu, yielding $M_S = 900$ emu/mol. In the case of sample WCN3#21 we can not evaluate $M_S$ in this way, because the *m(H)* curve has not been measured to high enough fields.

The diamagnetic background exhibited by the data in Fig. 4b), however, does not exactly follow a straight line. Therefore, the extrapolation back to *H*=0, to determine the magnetic moments to get the saturation magnetizations contains an uncertainty of at least 10%. The uncertainty is comparable for the evaluation of the magnetic moments at $H=1000$ Oe from the hysteresis curve. Therefore, we only give rounded values of the magnetizations evaluated.

In the following we discuss the conversion of the CGS emu units to the SI system [38] and the calculation of the magnetic moment per atom from our measurements. Since, 1 emu/mol = 1 (emu/cm$^3$)(cm$^3$/mol) with 1 emu/cm$^3$ equal to $10^3$ A/m in SI units, *i.e.*, 1 emu = $10^{-3}$ Am$^2$, it is 1 emu/mol equal to $10^{-3}$ (A/m) (m$^3$/mol). For a Cu$_{41}$Ni$_{59}$ alloy, 1 mol covers $V_{mol,CuNi} = 6.8 \times 10^{-6}$ m$^3$, yielding 1 emu/mol = $0.147 \times 10^3$ A/m for this alloy. To calculate the magnetic moment per atom of the alloy, $m_{at}$, we have to multiply this result for the magnetization (*i.e.* the magnetic moment per volume) with the volume occupied by one atom in the alloy, given by $V_{mol,CuNi}$ divided by the Avogadro number. Thus, $m_{at}$ in the presence of a magnetization of 1 emu/mol is $m_{at}$(1 emu/mol) = $0.166 \times 10^{-26}$ Am$^2$ = $0.179 \times 10^{-3}$ $\mu_B$. Here $\mu_B = e\hbar/2m_e = 0.9274 \times 10^{-23}$ Am$^2$ is the Bohr magneton, where $e$ is the elementary charge, $\hbar=h/2\pi$ with $h$ Planck's constant, and $m_e$ the electron mass. Thus, for $M_S$(WCN3#12)= 790 emu/mol we get $m_{at} = 0.14$ $\mu_B$, and $M_S$(WCN3#1)= 900 emu/mol yields $m_{at} = 0.16$ $\mu_B$, which both are equal or close to the bulk material value, respectively, for a Cu$_{41}$Ni$_{59}$ alloy of 0.14 $\mu_B$ [38]. The deviation of $m_{at}$ of sample #1 from the bulk material value for the Cu$_{41}$Ni$_{59}$ alloy can be explained by concentration variations. The Ni concentration varies up to $\pm 2$ at.% from its mean value of 58 at.% for samples of the WCN3 series investigated by RBS, and m$_{at}$ (sample #1) represents the Cu$_{40}$Ni$_{60}$ bulk material value.

Moreover, our $M_S$ values are comparable to those of Ruotolo *et al.* (Ref. [40], Fig. 4). They got for Cu$_{40}$Ni$_{60}$ multilayers with a thickness of the Cu$_{40}$Ni$_{60}$ layers from 4.2 nm to 34 nm, values for the saturation magnetization of $M_S$ = 72-112 emu/cm$^3$, resulting in (72-112)$\times 10^3$ A/m = (72-112)$\times 10^3$ / (0.147$\times 10^3$) emu/mol = (490-769) emu/mol, if we apply the conversion formulas of our Cu$_{41}$Ni$_{59}$ alloy. The highest value of 769 emu/mol may be compared with the result of our sample #1 ($d_{CuNi}$ = 48 nm). Our sample #12 ($d_{CuNi}$ = 21 nm) may be compared with the 17 nm sample of Ruotolo *et al.*, which has $M_S$=100 emu/cm$^3$, *i.e.* 680 emu/mol. In both cases our values are a factor of 1.2 larger than those of Ref. [40].

The saturation magnetization $M_S$ for our specimens obtained from the hysteresis measurements at $T = 2$K is $M_S$ = 790 emu/mol and 900 emu/mol for samples WCN3#12 and #1, respectively. In Fig. 3 the magnetization at T = 2 K of sample #21 is close to these values. For sample #12 the value of $M$ at the lowest temperature (T = 2 K) in Fig. 3 is somewhat smaller than $M_s$. For sample WCN3#1 it is much smaller than $M_s$. The reason is not yet clear.

Therefore, we evaluated the magnetic moment for samples #12 and #1 at H = 1000 Oe from the measured hysteresis curves. Considering the diamagnetic moment of the Si substrate, we got $2.0 \times 10^{-5}$ emu (#12) and $6.7 \times 10^{-5}$ emu (#1), respectively, resulting in a magnetization of 430 emu/mol (#12) and 550 emu/mol (#1). These values do not agree with those in Fig. 3 at the lowest temperature. For sample #12 the value of $M$ (1000 Oe at 2 K) obtained from the hysteresis is lower than in Fig. 3, whereas the opposite is the case for sample #1. This indicates that, the magnetic properties of our samples depend on the path on which the regarded state (e.g. at H = 1000 Oe and T = 2 K) is reached. The magnetic anisotropy of the film [40] seems to be not the only reason for this behavior which has to be studied in detail in a separate investigation.

## III. SUPERCONDUCTING PROPERTIES

### A. Single Nb layer

For material characterization we prepared single Nb layer samples of the structure Si(substrate)/Si(buffer layer)/Nb(flat)/Si(cap layer) with thickness of the buffer and protection layer of about 10 nm. Figure 5 shows the dependence of the superconducting $T_c$, determined by four-contact resistance measurements in a standard $^4$He cryostat, on the thickness of the Nb single layer. The behavior is similar to that one reported in Ref. [41]. The midpoint of the $R(T)$ curves at the superconducting transition was accounted for as transition temperature $T_c$. The width of transition ($0.1R_N$-$0.9R_N$ criteria, where $R_N$ is the normal state resistance just above $T_c$) for all investigated samples was not more than 0.1 K. The inset to Fig. 5 gives the cross-section TEM image of the layered structure from which one may deduce that the Nb layer is textured polycrystalline with very sharp interfaces to the surrounding silicon layers.

We measured the upper critical field $B_{c2\perp}$ perpendicular to the single Nb films to determine the superconducting coherence length $\xi_S$ of the Nb layer, which enters the proximity effect theory. First, the Ginzburg-Landau coherence length, $\xi_{GL}(0)$, is obtained from the linear dependence of $B_{c2\perp}(T)$ near the superconducting transition temperature making use of the expression [42]: $\xi_{GL}(0) = [-(dB_{c2\perp}(T)/dT)(2\pi T_{c0}/\Phi_0)]^{-1/2}$, where $\Phi_0 = 2.07 \times 10^{-15}$ T·m² is the magnetic flux quantum. For a "dirty" superconductor (short electron mean free path, $l_S \ll \xi_{BCS}$, with the Bardeen-Cooper-Schrieffer coherence length $\xi_{BCS} = \hbar v_S \gamma/(\pi^2 k_B T_{c0})$, where $\gamma \simeq 1.781$ is the Euler constant, $k_B$ is Boltzmann's constant, $v_S$ is the Fermi velocity of the superconductor, and $T_{c0}$ is the superconducting transition temperature of the Nb film at zero magnetic field) the coherence length $\xi_S$ is defined as follows [43,44]:

$$\xi_S = (\hbar D_S/2\pi k_B T_{c0})^{1/2} = \sqrt{(\pi/6\gamma)}\sqrt{(l_S\xi_{BCS})} = (2/\pi)\xi_{GL}(0), \quad (1)$$

where $D_S = l_S v_S/3$ is the electronic diffusion coefficient in a superconductor, with $l_S$ the electron mean free path, and $\xi_{GL}(0) = 0.855(l_S\xi_{BCS})^{1/2}$ for a dirty superconductor. Using the above formulas and our experimental data, yielding $-B_{c2\perp}(T)/dT = 0.558$ T/K for $d_{Nb} = 6.8$ nm and 0.372 T/K for $d_{Nb} = 14.0$ nm, we obtain $\xi_{GL}(0) = 9.7$ nm and 10.5 nm, *i.e.* $\xi_S = 6.16$ nm and 6.68 nm, respectively. We use the range of these values of $\xi_S$ as an initial guess when fitting the proximity theory to our experimental data in the present work.

### *B. Nb/Cu$_{41}$Ni$_{59}$ bilayers*

The resistance measurements of the Nb/Cu$_{41}$Ni$_{59}$ bilayers were performed in a $^3$He-cryostat and a dilution refrigerator. The standard DC four-probe method was used with a measuring current of 10 μA in the temperature range 0.4 K-10 K and of 2 μA for 40 mK-1.0 K, respectively. Possible thermoelectric voltages were eliminated by alternating the polarity of the current during the resistance measurements. The $T_c$ values were defined applying the same criteria as for the single Nb film. The shift between transitions measured for increasing and decreasing temperature was smaller than 15 mK.

Figure 6 demonstrates the dependence of the superconducting transition temperature on the thickness of the Cu$_{41}$Ni$_{59}$ layer for five values of the Nb layer thickness, where $d_{Nb}(S15) \approx 7.3$ nm, $d_{Nb}(S16) \approx 8.3$ nm, $d_{Nb}(S21) \approx 6.2$ nm, $d_{Nb}(S22) \approx 7.8$ nm, and $d_{Nb}(S23) \approx 14.1$ nm.

For specimens with $d_{Nb} \approx 14.1$ nm the transition temperature $T_c$ reveals a non-monotonic behavior with a very shallow minimum at about $d_{CuNi} \approx 6.8$ nm. It is just the qualitative behavior that has been observed in many other studies [25-34]. For $d_{Nb} \approx 8.3$ nm and 7.8 nm the minimum becomes clearly manifested. The transition temperature $T_c$ reveals an expressed non-monotonic behavior with a deep minimum at $d_{CuNi}$ about 7.0 nm and 7.9 nm, respectively. For the series of specimens with $d_{Nb} \approx 7.3$ nm the transition temperature $T_c$ decreases sharply for increasing ferromagnetic Cu$_{41}$Ni$_{59}$ layer thickness, until $d_{CuNi} \approx 3.8$ nm. Then, for $d_{CuNi} \approx 3.8$-12.8 nm, the superconducting transition temperature vanishes (at least $T_c < 40$ mK, which is the lowest temperature measured in our dilution refrigerator). For $d_{CuNi} > 12.8$ nm the transition into a superconducting state is observed again. Finally, $T_c$ increases to a little bit above 2 K. This phenomenon of the reentrant superconductivity in the S/F bilayer has been presented in our recent brief publication [24]. Meanwhile, we succeeded in the preparation of a sample with a even more thin Nb layer, $d_{Nb} \approx 6.2$ nm, showing an outstanding reentrant superconductivity behavior with evidence for a second disappearance of the superconducting state at $d_{CuNi} > 37.4$ nm. Altogether, the $T_c(d_{CuNi})$ curves given in Fig. 6 represent all types of non-monotonic $T_c(d_{CuNi})$ behaviors predicted by the theory [20,21].

Complimentary to the previous $T_c(d_{CuNi})$ dependences, data on $T_c(d_{Nb})$ with a $Cu_{41}Ni_{59}$ top layer of constant thickness are presented in Fig. 7. The optimal sample design implies a "thick" (which can be considered as infinitely thick in the theory) copper-nickel top layer to determine the critical thickness $d_{CuNi}^{cr}$, at which superconductivity vanishes. The critical thickness allows one to formulate a constraint (see below) which essentially restricts the freedom in varying the physical parameters of the theory. The relevant data are given for the samples with $d_{CuNi} \approx 56$ nm and $d_{CuNi} \approx 25$ nm, where the first sample series fulfills the requirement of "infinitely thick" F metal. The shadowed area indicates the values of $T_c(d_{CuNi} \to \infty)$ for Nb films of 6.2-8.0 nm thickness. This region of steep $T_c$ variation is the key condition to observe large-amplitude oscillations and reentrant behavior of the superconducting critical temperature of S/F bilayers as a function of the thickness $d_F$ of the F layer.

## IV. ANALYSIS OF THE DATA AND DISCUSSION

As it was mentioned in the Introduction, the proximity effect in the S/F system has a crucial difference from that of the superconductor/normal metal (S/N) system. In a usual S/N system the pairing function from the S-layer exponentially relaxes deep into the N-layer showing purely evanescent behavior on the scale of the coherence length [43], $\xi_N = (\hbar D_N/2\pi k_B T)^{1/2}$, where $D_N = l_N v_N/3$ is the diffusion coefficient in the normal metal, $v_N$ and $l_N$ are the Fermi velocity and the electron mean free path of the N metal, respectively.

In the F metal the singlet Cooper pairs are combined of electrons with opposite spin directions and with opposite directions of the wave number vectors, however, the absolute values of the wave number vectors are not equal because of the exchange splitting of the conduction band. This is the reason that the pairing occurs with a non-zero momentum given by $\hbar Q_{FM} = E_{ex}/v_F$. Here $E_{ex} \ll E_F$ is the exchange splitting energy of a free-electron-like, parabolic conduction band, $E_F$ is the Fermi energy, and $v_F$ is the Fermi velocity of the F metal. Contrary to the case of a non magnetic metal, the pairing function does not only decay in an F metal, but in addition oscillates on a wavelength scale $\lambda_{FM}$ (*i.e.* the wave number is $k_{FM} = 2\pi/\lambda_{FM}$) given by the magnetic coherence length $\xi_F$. For a clean ferromagnet ($l_F >> \xi_{F0}$) it is $\lambda_{FM}$ equal to $\lambda_{F0} \equiv 2\pi\xi_{F0} = 2\pi\hbar v_F/E_{ex}$ [45,46], whereas in the dirty case ($l_F << \xi_{F0}$) we have $\lambda_{FM}$ equal to $\lambda_{FD} = 2\pi\xi_{FD} = 2\pi(2\hbar D_F/E_{ex})^{1/2}$ [1,44]. Here, $D_F = l_F v_F/3$ with $l_F$ the electron mean free path in the F metal. The decay length of the pairing wave function is given by $l_F$ and $\xi_{FD}$ in the clean and dirty case, respectively [20,44,46]. In strong ferromagnets like Ni, Fe and Co the clean case oscillation length $\lambda_{F0}$ can be a few times shorter than the decay length, $l_F$, where $l_F \approx \min\{l_{\uparrow(\downarrow)}\}$, if $l_\uparrow >> l_\downarrow$ or vice versa. Here, $l_{\uparrow(\downarrow)}$ is the electron mean free path in the spin-up (spin-down) conduction subband. More accurately, $(l_F)^{-1} = \frac{1}{2}((l_\uparrow)^{-1}+(l_\downarrow)^{-1})$, where the shortest mean free path of the two ($l_\uparrow$, $l_\downarrow$) dominates in $(l_F)^{-1}$ whatever the spin-projection is (see details in Ref. [23]). In a dirty ferromagnetic metal both characteristic lengths are equal [44] (more accurately, the statement is correct if the conduction band exchange splitting energy is still much larger than the thermal energy, see Ref. [47]).

The acquisition of phase by the pairing wave function results in interference phenomena at S/F interfaces of a layered system. The interference conditions change periodically between constructive and destructive interference when the ferromagnetic layer thickness changes. Then, the pairing function flux through the S/F interface becomes periodically modulated as a function of the ferromagnetic layer thickness $d_F$. As a result, the coupling between the S and F layers is modulated, and $T_c$ oscillates as a function of $d_F$ [20,35]. The amplitude of the $T_c$ oscillation depends sensitively on the superconducting layer thickness.

As we mentioned in our previous paper [24] we failed to reproduce the non-monotonous $T_c(d_{CuNi})$ behavior, neither using the dirty-case, Usadel theory [48] with "2/5" correction factor in the diffusion coefficient [49,50], nor the multi-mode solution to the dirty-case proximity effect equations [51, 52]. A general, Eilenberger-type theory [23] could be more suitable to describe the experiments, however, the intermediate case ($l_F \approx \xi_{F0}$, i.e. the crossover region between the dirty and the clean cases) represented by our samples is the case hardly accessible for a quantitative description. The only theory that appears to give an approximate but consistent description of the data is our extension of the dirty-case theory towards the clean case, beyond the validity of the basic inequality $l_F << \xi_{F0}$, applied on the intermediate region [20, 35] (for details see the Appendix of the present work).

Here, we only recall that five physical parameters enter the theory: (1) the superconducting coherence length $\xi_S$, (2) the magnetic coherence length $\xi_{F0}$, (3) the mean free path of conduction electrons in a ferromagnet $l_F$, (4) the ratio of Sharvin conductances at the S/F interface $N_Fv_F/N_Sv_S$, and (5) the interface transparency parameter $T_F$. As it was described above, the initial guess value for the superconducting coherence length $\xi_S$ is obtained from the upper critical field measurements. Availability of the critical thickness $d_{Nb}^{cr}(d_{CuNi}\rightarrow\infty) \approx 6.0$ nm allows us to impose a constraint on the parameters $N_Fv_F/N_Sv_S$ and $T_F$ using the expression [35,53],

$$d_{Nb}^{cr} = \xi_S\sqrt{2\gamma}\arctan\left\{\frac{\pi}{\sqrt{2\gamma}}\left(\frac{\xi_{BCS}}{\xi_S}\right)\frac{N_Fv_F/N_Sv_S}{1+2/T_F}\right\}, \qquad (2)$$

as follows

$$\frac{N_Fv_F}{N_Sv_S}\frac{1}{1+2/T_F} \approx 0.047, \qquad (3)$$

where the values of parameters $\xi_S = 6.2$ nm, $\xi_{BCS} = 42$ nm, valid for Nb with a reduced critical temperature of $T_{c0} = 8.75$K, compared to bulk material, as obtained for a high quality thin film of 19 nm thickness in Ref. [11], and $d_{Nb}^{cr} \approx 6.0$ nm have been used. Now, $N_Fv_F/N_Sv_S$, as well as $T_F$, can be varied, but their values must be kept confined according to Eq. (3). With the three parameters (of the five) constrained by the experimental data, the problem of consistent fitting the six curves from Figs. 6 and 7 becomes feasible.

The results of the fitting are plotted in Figs. 6 and 7 by solid lines. The fitting looks not completely perfect, but quite reasonable, because all specific features of the $T_c(d_{CuNi})$ and $T_c(d_{Nb})$ behavior are reproduced correctly for the used sets of the physical parameters. We varied a little the superconducting coherence length $\xi_S$ in accordance with Eq. (1), because we expect a little decrease of the mean-free path (and consequently the diffusion coefficient $D_S$) upon decreasing the Nb film thickness.

The curves in Fig. 6 were calculated using the following parameters for curves S15, S16, S21, S22, and S23, with $T_{c0,Nb}(d_{CuNi} = 0nm) = 6.67, 7.0, 6.2, 6.85,$ and $8.0$K, respectively (taken from the fit in Fig. 5): $\xi_S = 6.3, 6.4, 6.1, 6.5,$ and $6.6$ nm; $N_Fv_F/N_Sv_S = 0.22$ for all; $T_F = 0.67, 0.66, 0.65, 0.61,$ and $0.44$; $l_F/\xi_{F0} = 1.3, 1.3, 1.1, 1.1,$ and $1.1$; $\xi_{F0} = 9.5, 9.5, 11.2, 10.7,$ and $10.8$ nm. The curve in Fig. 7 was calculated using $T_{c0,Nb}=9.1$ K, $\xi_S = 6.1$ nm, $N_Fv_F/N_Sv_S = 0.22$, $T_F = 0.61$, $l_F/\xi_{F0} = 1.1$, $d_F/\xi_{F0} = 50.0$, where $\xi_{F0} = 10.5$ nm. We did not vary the superconducting coherence length $\xi_S(l_s,T_{c0})$ and also not $\xi_{BCS}(T_{c0})$ when the niobium layer thickness was varied upon calculating $T_c(d_{Nb})$, because from our experience in such type of corrections (see Fig. 6 of our paper [35], where we considered changes of $\xi_S$ and $\xi_{BCS}$ with the critical temperature for a free standing Nb Film) we do not expect that consistency of the physical picture will be broken. The electron mean free path $l_F \approx 11.8 – 12.4$ nm, used in our calculations, appeared to be longer than the coherence length $\xi_{F0} = 9.5 – 11.2$ nm. According to Ref. [31], $l_F \approx 4.4$ nm for a $Cu_{47}Ni_{53}$ alloy with resistivity $\rho_F = 57$ μΩ·cm (bulk material, $T = 10$K). Assuming that the product $<\rho_F l_F> \approx 2.5 \cdot 10^{-5}$ μΩ·cm$^2$ remains unchanged upon adding impurities [54] we get $l_F \sim 10$ nm for our $Cu_{41}Ni_{59}$ alloy using our data for the low-temperature resistivity [24], $\rho_F \approx 25$ μΩ·cm. Thus, both the proximity and the resistivity analysis indicate that our $Cu_{41}Ni_{59}$ alloy is between the "dirty" ($l_F << \xi_{F0}$) and the "clean" ($l_F >> \xi_{F0}$) cases.

So far, our experimental data on S/F bilayers (with F = $Cu_{41}Ni_{59}$ alloy) were fitted by theoretical curves, using a value for $\xi_S$ of about 10 nm (see Refs. [24] and [55,56] for $T_c(d_F)$ dependencies of sample series S15, S16 and of S21, S22, S23, respectively, and Refs. [24,56,57] for $T_c(d_S)$ curves), instead of the improved value in the range of $\xi_S = 6.2$-$6.7$ nm obtained from critical field measurements (see Sec. III B). Also for $\xi_S$ about 10 nm a set of parameters could be found to fit the data in Figs. 6 and 7. In this case $\xi_S = 9.6 – 11.0$ nm, $N_F v_F / N_S v_S = 0.17 – 0.23$, $T_F = 0.43$-$0.85$, $l_F / \xi_{F0} = 1.1 – 1.2$, and $\xi_{F0} = 8.6 – 11.0$ nm. These values are close to the set of parameters given above and, thus, show that the phenomenon does not so strongly depend on $\xi_S$.

Finally, using material parameters obtained from the non-monotonous and reentrant superconductivity behavior, we succeeded to observe in Nb/$Cu_{41}Ni_{59}$ bilayers, we can plot $T_c(d_F)$ curves of a F/S/F spin-switch core structure [17,20] and estimate $\Delta T_c = T_c^{AP} - T_c^{P}$, where the AP superscript stands for the antiparallel alignment, and P for the parallel alignment of magnetizations in the F/S/F trilayer. For the derivation details see the Appendix. The results of such calculations are plotted in Fig. 8 (the values of the parameters are given in the figure caption). They show that for a Nb layer thickness in the range, $d_{Nb} \approx (2*6.0 – 2*7.5)$ nm = $(12.0 – 15.0)$ nm, $\Delta T_c$ can be as large as 2 K. In Fig. 9 we plotted the maximal change of the critical temperature, $\Delta T_c^{max} = \max(T_c^{AP} - T_c^{P})$ due to a change of the magnetization of the CuNi layers from parallel to antiparallel, together with the related thickness $d_{CuNi}$ as a function of the Nb layer thickness. The value of $\Delta T_c^{max}$ increases for decreasing Nb layer thickness. The related (*i.e.* "optimized") CuNi layer thickness passes through a maximum and then decreases. The shaded region should be available for our experimental parameters range. We believe that failure to observe a large spin-switch effect in [26,30,58-68] is because of non-optimal choice of materials and layer thicknesses.

## V. CONCLUSIONS

In the present work the complete set of non monotonic behavior of the critical temperature on the thickness $d_F$ of the F metal in a S/F bilayer predicted by theory could be realized experimentally, using Nb as a superconductor and $Cu_{41}Ni_{59}$ alloy as a ferromagnetic metal.

The effect depends strongly on the thickness $d_S$ of the superconducting layer. The value of $d_S$ has to be chosen in a certain range above the critical thickness (for which superconductivity vanishes in a S/F bilayer with "infinite" thickness of the F metal). Only here the system reacts sensitively enough on changes of $d_F$. For sample series with fixed $d_S$ above this range (*e.g.* for $d_S$ = 14.1 nm), the $T_c(d_F)$ dependence only shows a slight suppression with a shallow minimum. A reduction of $d_S$ (to 8.3 nm and 7.8 nm) yields $T_c(d_F)$ curves with a strong suppression of superconductivity and a deep minimum. For even lower values of $d_S$ (7.3 nm and 6.2 nm) superconductivity vanishes for a certain range of $d_F$ and then restores again, *i.e.* reentrant superconducting behavior is observed.

The extinction region of superconductivity is especially broad for the sample with the thinnest S layer. In this case a second extinction of superconductivity is observed. Thus, evidence for the multiple reentrant behavior predicted by theory has been found. The theoretical curve in that case does not give a further reentrance of the superconducting state for higher $d_F$ values. However, for a slightly thicker Nb layer ($d_{Nb} \approx 6.3$ nm) the next island of superconductivity is expected above $d_{CuNi} \approx 51$ nm. It will be searched for the second reentrance of superconductivity in further investigations.

Our experiments clearly demonstrate the existence of a quasi-one-dimensional FFLO like state in our S/F bilayers. In this state the non monotonic behavior of the critical temperature, as well as the extinction and recovery of superconductivity are governed by interference effects of the superconducting pairing wave function. The situation is similar to the optical analogue of interference of light in a parallel sided plate of glass with a mirror coated back side at normal incidence, in which the interference conditions change between constructive and destructive when changing the thickness of the plate.

To demonstrate the predicted effect clearly in experiment, several special techniques were joined together: All samples of a series were fabricated in the same run using a wedge technique. The Nb target was moved during full power sputtering, yielding a high quality flat S layer with small thickness. RBS was applied for a precise determination of the thickness and alloy composition of the thin metal films of the S/F bilayers. A "window" for $d_S$ close to the critical thickness was chosen. The high quality of the samples was demonstrated by cross-sectional TEM investigations, scanning Auger electron spectroscopy and SQUID magnetometry.

All experimental curves could be described by the theory, using a suitable set of fitting parameters. An improved value of the superconducting coherence length $\xi_S$ has been used. It turns, however, out that the theoretical curves do not strongly depend on the value of $\xi_S$. This is an important information for the fabrication of the F/S/F superconducting spin switch in which the thickness of the S layer has to be chosen twice compared to the case of a S/F bilayer (which will probably increase the electron mean free path $l_S$ and, thus, the value of $\xi_S$) to receive a similar $T_c(d_F)$ behavior.

In the superconducting F/S/F spin switch, the critical temperature $T_c$, however, depends on the relative direction of the magnetizations of the F layers. The $T_c$ is lower in the parallel case compared to the antiparallel case. For specimens close to the extinction region of a sample series showing reentrant superconductivity, the difference can be several Kelvin, as we calculated for a fictive sample using our experimental parameters. A systematic study of F/S/F trilayers and the realization of the F/S/F spin switch will be our next experimental tasks.

## VI. ACKNOWLEDGMENTS

The authors are grateful to J. Aarts, C. Attanasio, A. Golubov, M. Kuprianov, and V. Oboznov for stimulating discussions, and R. Horny, D. Vieweg, S. Heidemeyer, B. Knoblich and W. Reiber for assistance in measurements. The work was supported in parts by Russian Fund for Basic Research (RFBR) under the grants Nos 07-02-00963-a, 08-02-90105_Mol_a, and 09-02-12260-ofi_m. Furthermore, the project was supported by the Deutsche Forschungsgemeinschaft (DFG) under the grant „Study of the Superconducting Proximity Effect Spin-Valve Phenomenon in Superconductor/Ferromagnet Nanolayered Structures" (GZ: HO 955/6-1).

## VII. APPENDIX

In the Appendix we derive expressions for calculating the superconducting $T_c$ of a F/S/F spin-switch core structure with physical parameters obtained from our experiments on S/F bilayers. The formulas for $T_c$ of S/F bilayers used for calculations of the curves in Figs. 6 and 7 follow as a particular case of the F/S/F structure with parallel alignment of the F-layer magnetizations.

Before we proceed with derivations let us note that our estimations made from the proximity effect as well as the resistivity data indicate that the coherence length $\xi_{F0}$ and the conduction electron mean free path $l_F$ are of the same order, $\xi_{F0} \sim l_F \sim 10$ nm. So our samples refer to the intermediate case in between of the dirty and the clean cases. Strictly speaking, the dirty case theory based on the Usadel equations [48] is valid at the condition $l_F \ll \xi_{F0}$ which is clearly not fulfilled in our samples. Then, the Eilenberger theory [69], reformulated for S/F hybrids [23,70,71], must came into play. The advantage of the Eilenberger formulation is that it can be applied for arbitrary electron mean-free path, however, the equations are anisotropic and hard to solve analytically. A solution for the strong and clean enough ferromagnet ($\xi_{F0} \ll l_F$) was proposed in Ref. [23], while analysis for the weak proximity-effect regime (low-transparent S/F interface) and arbitrary $\xi_{F0}$ and $l_F$ was given in a recent paper [71]. The calculations of superconducting $T_c$ versus $d_F$ [23], and the spatially resolved density of states (DOS) in the ferromagnet ([71], Fig. 4b), both for S/F bilayers and weak impurity scattering, have shown slowly decaying oscillations never observed experimentally [1,2]. Because of the inhomogeneous nature of the boundary problem, the Anderson theorem (i.e. the insensitivity of the superconducting state to the presence of nonmagnetic impurities) does not work anymore, and non-magnetic impurities as well as magnetic impurities and spin-orbit scatterings give rise to the damping of the pairing function oscillations [20,71-73], though the scattering terms enter the Eilenberger equations in a different way. It is important that magnetically active scatterings and interfaces can couple the singlet and triplet superconducting pairing components of different symmetries [52,71,74,75]. In the particular case of stoichiometrically disordered ferromagnetic alloys, like $Cu_{1-x}Ni_x$ or $Pd_{1-x}Fe_x$, an electron scattering at microscopic inhomogeneities of the alloy composition may dominate. All these circumstances greatly complicate an accurate solution of the problem even for S/F bilayers.

A palliative solution to the problem was proposed in Ref. [20], where the anisotropic Eilenberger kernel was numerically averaged over trajectories, and the resulting oscillating decay was fitted with a single mode of the complex wave vector, varying the ratio $l_F/\xi_{F0}$ in the range 0.5-5 (see pages 155-156 in Ref. [20]). It appeared that the mode wave vector fits the Usadel's solution wave vector with the replacement of $D_F = l_F v_F/3$ by $D_F = v_F l_F/(1+i l_F/\xi_{F0})$. The same result was obtained by Linder *et al.* [71] when they considered the dirty limit of their general, Eilenberger equations solutions (see formulas (22) and (23) of Ref. [71]). Both papers notice explicitly that the diffusion coefficient in one dimension, $D_F = v_F l_F$, stands instead of the three-dimensional diffusion coefficient $D_F = v_F l_F/3$ in the above expression (see the sentence just after formula (23) in Ref. [71], and the first paragraph in page 156 of Ref. [20] with an obvious symbol conversion typo in line 3, to be corrected as follows: $\leq \rightarrow$ \simeq). One has to realize clearly that the single decay length that appears in the extended Usadel approach absorbs approximately all types of electron scatterings including the spin-reversal scatterings too.

We consider a $F_L/S/F_R$ trilayer with the S layer thickness $d_S = 2\tilde{d}_S$, because the spin-switch core structure can be regarded as a stack of two bilayers $F/\tilde{S}$ and $\tilde{S}/F$, forming a $F/2\tilde{S}/F$ trilayer with the left F layer thickness $d_{FL}$, and the right F layer thickness $d_{FR}$. The magnetization directions of the ferromagnetic $F_L$ and $F_R$ layers are considered to be either parallel (P alignment) or antiparallel (AP alignment). Other magnetic parameters as well as the layers thickness and transparencies of the $F_L/S$ and $S/F_R$ interfaces can be different. This is because the growth conditions for the first (*i.e.* $F_L$ on a substrate or an exchange bias layer) and the second ($F_R$ on the S layer) ferromagnetic layers are essentially different. Using our advanced wedge deposition technique we can diminish the influence of possible scatter in the deposition conditions, preparing the bottom or top (or the both if necessary) ferromagnetic layers as wedges, and searching for the optimal specimen after cutting the long wedge-shape sample into strips.

A similar nonsymmetrical structure was examined recently by Fauré *et al.* [76] (and also by Cadden-Zimansky *et al.* [77] with different F-layer thickness but the other parameters being identical), however, they considered a thin superconducting layer, $2\tilde{d}_S \leq \xi_S$ (notice here that in the notations of paper [76] the superconducting layer thickness is $d_S \leq \xi_S$). Our analysis is valid also at the condition $2\tilde{d}_S > \xi_S$, which follows from our experimental data on bilayers.

To solve the problem of finding the critical temperature $T_c$ (either $T_c^P$ for the P alignment, or $T_c^{AP}$ for the AP alignment) we solve the linearized Usadel equations [1,48] for the pairing function $\Phi(x,\omega > 0)$:

$$\left\{ \omega + iE_{ex}/2 - \tfrac{1}{2} D_F \frac{d^2}{dx^2} \right\} \Phi_F(x,\omega) = 0 \qquad (A1)$$

for each of the F layers, and

$$\left\{ \omega - \tfrac{1}{2} D_S \frac{d^2}{dx^2} \right\} \Phi_S(x,\omega) = \Delta(x) \qquad (A2)$$

for the S layer, where $\Delta(x)$ is the superconducting order parameter, $D_S(D_F)$ is the diffusion coefficient of electrons in the S(F) layer, $E_{ex}$ is the exchange splitting of the conduction band, and $\omega = \pi T(2n+1)$ is the Matsubara frequency. Here, we set $\hbar$ and $k_B$ equal to unity. The solutions have to satisfy the boundary conditions [20,46,78]

$$\frac{d}{dx}\Phi_F(\pm \tilde{d}_S \pm d_{\substack{FR \\ FL}}, \omega) = 0 \qquad (A3)$$

at outer surfaces, and

$$N_S D_S \frac{d}{dx}\Phi_S = N_F D_F \frac{d}{dx}\Phi_F, \qquad (A4)$$

$$\mp D_F(\mathbf{n}_F \cdot \nabla_x \Phi_F) = \frac{v_F T_F}{2}(\Phi_S - \Phi_F) \qquad (A5)$$

at the inner interfaces of the superconductor with the ferromagnets. In equations (A3)-(A5), $N_S(N_F)$ is the electronic density of states of the S(F) layer, $\mathbf{n}_F$ is a vector of the outward unit normal to the right (–) or left (+) S/F interface, $T_F$ is the dimensionless interface transparency parameter, $T_F \in [0,\infty)$ [20,78], $v_F$ is the Fermi velocity of the ferromagnetic alloy, and x the space coordinate.

The reduced critical temperature $t_c = T_c/T_{c0}$ in the single-mode approximation is found by solving the equation

$$\ln t_c = \Psi\left(\frac{1}{2}\right) - \mathrm{Re}\,\Psi\left(\frac{1}{2} + \frac{\phi^2}{2t_c(\tilde{d}_S/\xi_S)^2}\right), \qquad (A6)$$

where $\phi = k_S \tilde{d}_S$, and $k_S$ is the propagation momentum of the pairing function in the S layer, $\phi = \phi^P$ for the P alignment of magnetizations, and $\phi = \phi^{AP}$ for the AP alignment of magnetizations, $\xi_S = (D_S/2\pi T_{c0})^{1/2}$ is the superconducting coherence length in the S layer, $T_{c0}$ is the critical temperature of the stand-alone superconducting layer.

Consider first the P alignment of the magnetizations in the F layers. Matching solutions of the equations (A1) and (A2) at the S/$F_R$ interface we get the equation for $\phi^P$:

$$\phi^P \frac{c_1 \tan \phi^P - c_2}{c_1 + c_2 \tan \phi^P} = R_R = \frac{N_{FR} D_{FR}}{N_S D_S} \frac{k_{FR} d_S \tanh(k_{FR} d_{FR})}{1 + \frac{2 D_{FR} k_{FR}}{T_{FR} v_{FR}} \tanh(k_{FR} d_{FR})}. \qquad (A7)$$

A similar procedure for the $F_L$/S interface gives

$$\phi^P \frac{c_1 \tan \phi^P + c_2}{c_1 - c_2 \tan \phi^P} = R_L = \frac{N_{FL} D_{FL}}{N_S D_S} \frac{k_{FL} d_S \tanh(k_{FL} d_{FL})}{1 + \frac{2 D_{FL} k_{FL}}{T_{FL} v_{FL}} \tanh(k_{FL} d_{FL})}, \qquad (A8)$$

where the coefficients $c_1$ and $c_2$ of the Usadel equation solutions in the S layer can be eliminated, thus giving the closed equation for finding $\phi^P$ at the P alignment:

$$\left(\phi^P \tan\phi^P - R\right)\left(R \tan\phi^P + \phi^P\right) + \Delta^2 \tan\phi^P = 0, \tag{A9}$$

where

$$R = \tfrac{1}{2}(R_L + R_R), \text{ and } \Delta = \tfrac{1}{2}(R_L - R_R). \tag{A10}$$

When inserted into equation (A6) the solution of (A9) for $\phi^P$ gives the critical temperature $T_c^P$ of the non-symmetric $F_L$/S/$F_R$ structure for the P alignment of magnetizations. In equations (A7) and (A8) $k_F$ is the complex-valued propagation momentum of the pairing function in the F layers.

For the AP alignment (the $F_R$ layer magnetization is reversed) the right-hand side of the matching condition (A7) changes to the complex-conjugated one: $R_R \to R_R^*$. Then, the equation for finding $\phi^{AP}$ for the AP alignment reads:

$$\left(\phi^{AP} \tan\phi^{AP} - \hat{R}\right)\left(\hat{R} \tan\phi^{AP} + \phi^{AP}\right) + \hat{\Delta}^2 \tan\phi^{AP} = 0, \tag{A11}$$

where

$$\hat{R} = \tfrac{1}{2}(R_L + R_R^*), \text{ and } \hat{\Delta} = \tfrac{1}{2}(R_L - R_R^*). \tag{A12}$$

The solution of equations (A11) and (A6) provides the critical temperature $T_c^{AP}$ of the $F_L$/S/$F_R$ structure at the AP alignment of magnetizations.

Now we consider in detail the symmetric spin-switch core structure with equivalent physical parameters of the ferromagnetic layers and interfaces. Then, $R_L = R_R = R$, and equation (A9) for finding $\phi^P$ for the P alignment of magnetizations reduces to

$$\phi^P \tan\phi^P - R = 0 \tag{A13}$$

where $R$ is given by Eq. (A7). Equation (A13) is just the equation for the S/F bilayer with the S-layer thickness equal to $\tilde{d}_S$ [20,35]. As we mentioned above, the trilayer with the S-layer thickness $2\tilde{d}_S$ and parallel alignment of magnetizations can be viewed as a symmetric stack of two bilayers, F/$\tilde{S}$ and $\tilde{S}$/F, with the $\tilde{S}$-layer thickness $\tilde{d}_S$ for each: the superconducting critical temperatures of the both systems are equal. If so, it is convenient here to parametrize equation (A13) as it is used for the analysis of the experimental data for the S/F bilayers in the present work.

The constituents of equation (A7) can be parametrized as follows:

$$k_F d_F = \sqrt{\frac{iE_{ex}}{D_F}} d_F = \frac{d_F}{\xi_{F0}} \sqrt{i\frac{\xi_{F0}}{l_F} - 1}, \quad (A14)$$

where $\xi_{F0} = v_F / E_{ex}$, and $l_F$ are the coherence length and the electron mean free path determined in Section IV, $D_F = v_F l_F / (1 + i l_F / \xi_{F0})$. Next,

$$D_F k_F / v_F = \left(1 - i\frac{\xi_{F0}}{l_F}\right)^{-1/2}, \quad (A15)$$

and, finally,

$$\frac{d_S}{D_S} = \frac{\pi}{2\gamma} \frac{d_S}{v_S} \frac{\xi_{BCS}}{\xi_S^2}, \quad (A16)$$

where the following definitions had been used:

$$\xi_S^2 = \frac{D_S}{2\pi T_{C0}}, \quad \xi_{BCS} = \frac{\gamma v_S}{\pi^2 T_{C0}}, \quad (A17)$$

$\gamma \approx 1.781$ is the Euler constant. Substitution of Eqs. (A14)-(A16) into Eq. (A7) yields

$$R = \frac{\pi}{2\gamma} \frac{N_F v_F}{N_S v_S} \frac{d_S \xi_{BCS}}{\xi_S^2} \frac{\tanh\left(\frac{d_F}{\xi_{F0}} \sqrt{i\xi_{F0}/l_F - 1}\right)}{\sqrt{1 - i\xi_{F0}/l_F} + (2/T_F)\tanh\left(\frac{d_F}{\xi_{F0}} \sqrt{i\xi_{F0}/l_F - 1}\right)}, \quad (A18)$$

which was actually used in combination with Eqs. (A13) and (A6) to calculate $T_c(d_F)$ and $T_c(d_S)$ curves in Figs. 6 and 7, respectively.

For the symmetric spin-switch core structure equation (A11) for the AP alignment of magnetizations reads [17],

$$(\phi^{AP} \tan\phi^{AP} - R')(R'\tan\phi^{AP} + \phi^{AP}) + (R'')^2 \tan\phi^{AP} = 0, \quad (A19)$$

where, according to Eq. (A12), $R'$ and $R''$ are the real and imaginary parts of $R$ given by Eq. (A18). Now, equations (A19), (A18) and (A6) solve the problem of finding the superconducting $T_c$ in the symmetric F/S/F structure for the antiparallel alignment of magnetizations of the F layers.

Having values of the physical parameters obtained from fitting of $T_c(d_F)$ and $T_c(d_S)$ for S/F bilayers we can estimate the magnitude of the spin-switch effect expected for the symmetric F/S/F core structure made of materials studied in this work. Figure 8 displays the results of $T_c(d_F)$ calculations for the F/S/F structure at P and AP alignment of magnetizations of the ferromagnetic layers.

**Figure captions**

Fig. 1 (color online). (a) Thickness of the Nb and $Cu_{1-x}Ni_x$ layers and Ni content in the $Cu_{1-x}Ni_x$ alloy, measured by RBS. The inset shows a sketch of the layers stack. Black square symbols for the thickness of the $Cu_{1-x}Ni_x$ alloy layer are measured points, whereas gray (orange) symbols were linearly interpolated. (b) Transmission electron microscopy (TEM) cross-sectional image of a cut across the layers of sample S22#18 marked by a red circle in the RBS data. According to RBS it is $d_{Nb} \approx 7.2$ nm and $d_{CuNi} \approx 14.1$ nm.

Fig. 2 (color online). (a) Scanning Auger electron spectroscopy (AES) of a Nb film grown on Si(wafer)/Si(buffer) substrate and capped by Si, of the sample with $d_{Nb} \approx 9.3$ nm of Fig. 5. The top panel shows the topographic view of the crater scarp. The bottom panel shows the sputter depth profiling of elements. (b) Scanning AES of a Si(substrate)/Si(buffer)/Nb/$Cu_{1-x}Ni_x$/Si(cap) sample. Series S22 sample #7, $d_{Nb}$ =7.5 nm and $d_{CuNi}$=32.9 nm according to RBS (linear interpolation between samples #6 and #9).

Fig. 3 (color online). Temperature dependence of the magnetization $M$ (measured in a field of $H$=1000 Oe) of the $Cu_{41}Ni_{59}$ wedge WCN3. Samples: #21 – $d_{CuNi} \approx 8$ nm, #12 – $d_{CuNi} \approx 21$ nm, and #1 – $d_{CuNi} \approx 48$ nm, according to RBS. Sample #21 and #1 measured upon cooling and sample #12 on warming. Zero point of $M$ at 200K slightly shifted so that the slope of the curves can be more clearly recognized.

Fig. 4 (color online) a) Hysteresis loop of the magnetic moment $m(H)$ of sample #12 of the $Cu_{41}Ni_{59}$ wedge WCN3, with $d_{CuNi} \approx 21$ nm (according to RBS determination) measured at 2 K. b) Hysteresis curve for a larger range of the magnetic field. Straight line serves to extrapolate $m_S$ (for details see the text).

Fig. 5 (color online). Dependence of the superconducting critical temperature of single Nb films on their thickness. The solid line is the result of fitting to the dependence $T_c = T_{c0}\{1-d_1/d_S+(d_2/d_S)^2\}$ with $T_{c0} = 9.3$K, $d_1 = 2.07$ nm, $d_2 = 4\times10^{-5}$ nm. The inset shows the TEM cross-section image of the sample with $d_{Nb} \approx 6.8$ nm.

Fig. 6 (color online). Non-monotonous $T_c$ ($d_F$) dependence for Nb/Cu$_{41}$Ni$_{59}$ bilayers (in the sequence of increasing S layer thickness): (a) S15 – $d_{Nb}\approx$7.3 nm and S16 – $d_{Nb}\approx$8.3 nm; (b) S21 – $d_{Nb}\approx$6.2 nm, S22 – $d_{Nb}\approx$7.8 nm, S23 – $d_{Nb}\approx$14.1 nm (see Fig. 1 for RBS data on case S22). Transition widths are within the point size if error bars not visible. Solid lines are fits from the theory (see the text).

Fig. 7 (color online). The $T_c(d_{Nb})$ dependence for a sample with a Cu$_{41}$Ni$_{59}$ top layer of constant thickness and varying Nb layer thickness. The calculated critical thickness ($T_c\rightarrow 0$ K) is $d_{Nb}^{cr} \approx 6.0$ nm. Solid line is a fit from the theory calculated with the parameters given in the text. The range of the Nb layer thickness $d_{Nb}\approx$6.2 – 8.0 nm most sensitive to $d_{CuNi}$ variations is shaded.

Fig. 8 (color online). The $T_c(d_F)$ curves of a hypothetical superconducting F/S/F spin-switch core structure, with $d_S = d_{Nb}$= 12.0 nm (a), $d_S = d_{Nb}$= 13.9 nm (b), and $d_S = d_{Nb}$= 15.0 nm (c), calculated using the following set of parameters for (a), (b), and (c) respectively: $T_{c0,Nb}(d_{CuNi} = 0$ nm$) = $ 7.7, 7.9, 8.15 K; taken in accordance with Fig. 5, and in all cases $\xi_S$ = 6.6 nm; $N_F v_F/N_S v_S$ = 0.22; $T_F$ = 0.6; $l_F/\xi_{F0}$ = 1.1; $\xi_{F0}$ = 10.5 nm.

Fig. 9 (color online). The maximal difference of critical temperatures for the AP and P alignments of magnetizations in the symmetric F/S/F spin-switch core structure, $\Delta T_c^{max} = \max\left(T_c^{AP} - T_c^{P}\right)$, versus S (*i.e.* Nb) layer thickness is shown by the solid curve. The values of parameters for the calculations are the same as in Fig. 8. The thickness of the F (*i.e.* CuNi) layers, at which an actual maximum is reached (see Fig. 8), is given by the dashed curve. The shaded region allows to select a range of the Nb layer thickness ($d_{Nb}$=12.5-13.8 nm) and the CuNi layer thickness ($d_{CuNi}$=3.5-5.0 nm) for which $\Delta T_c^{max}$ is in the range 1-2K.

**Figures**

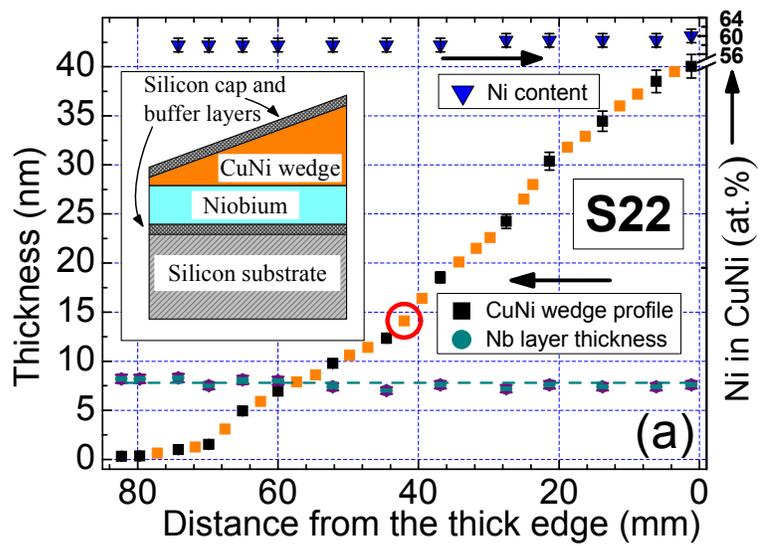

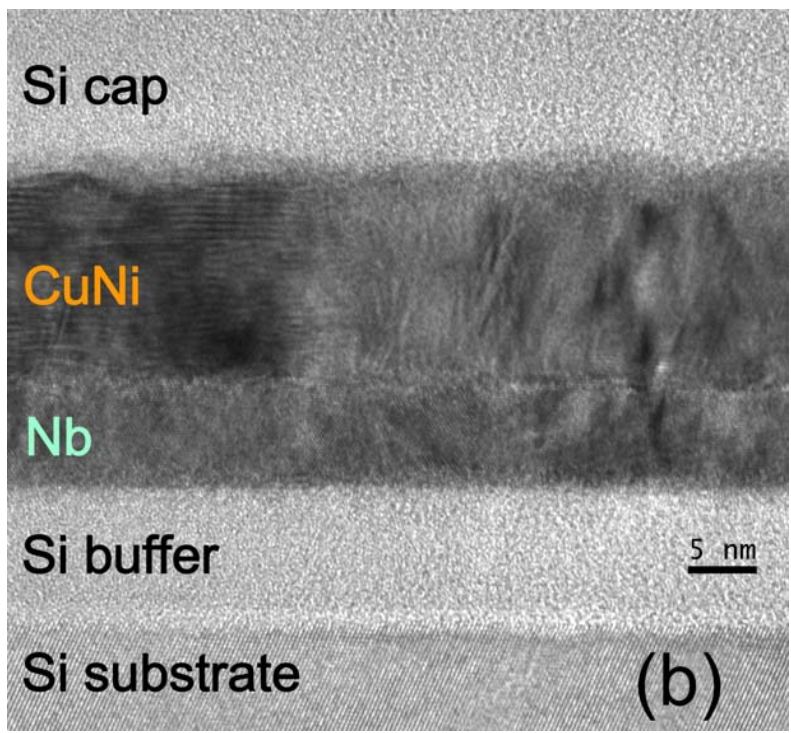

Fig.1a (top) and Fig.1b (bottom)

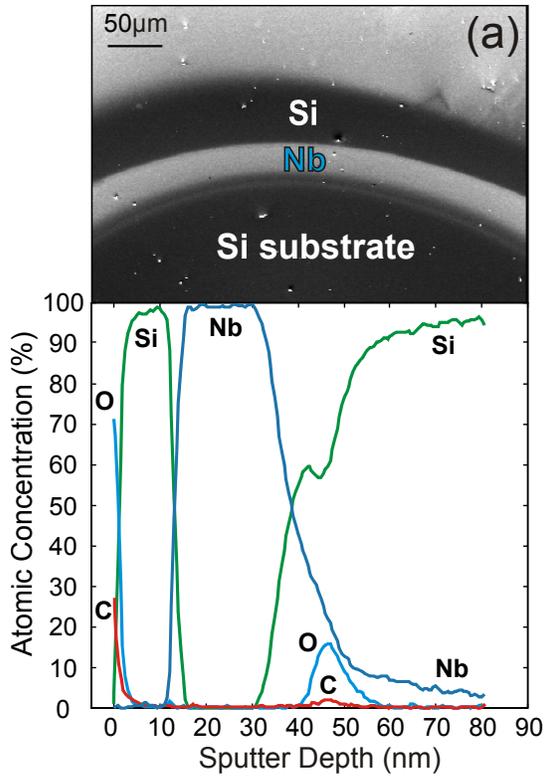

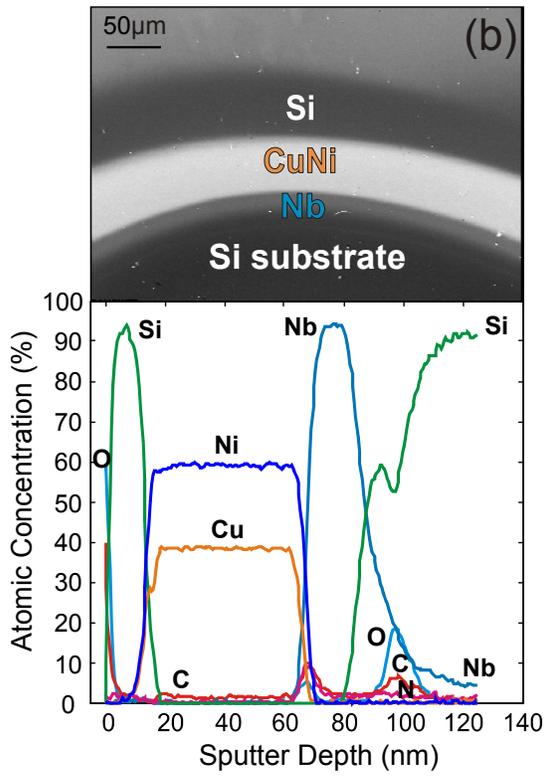

Fig.2a (top) and Fig.2b (bottom)

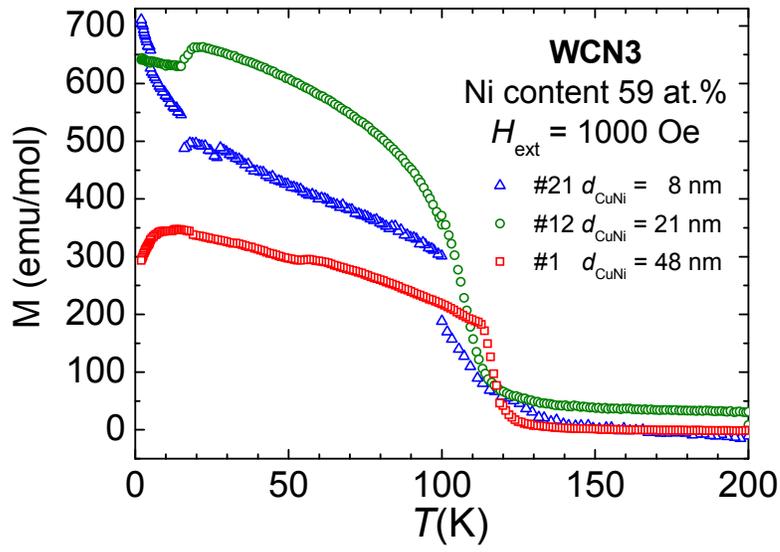

Fig.3

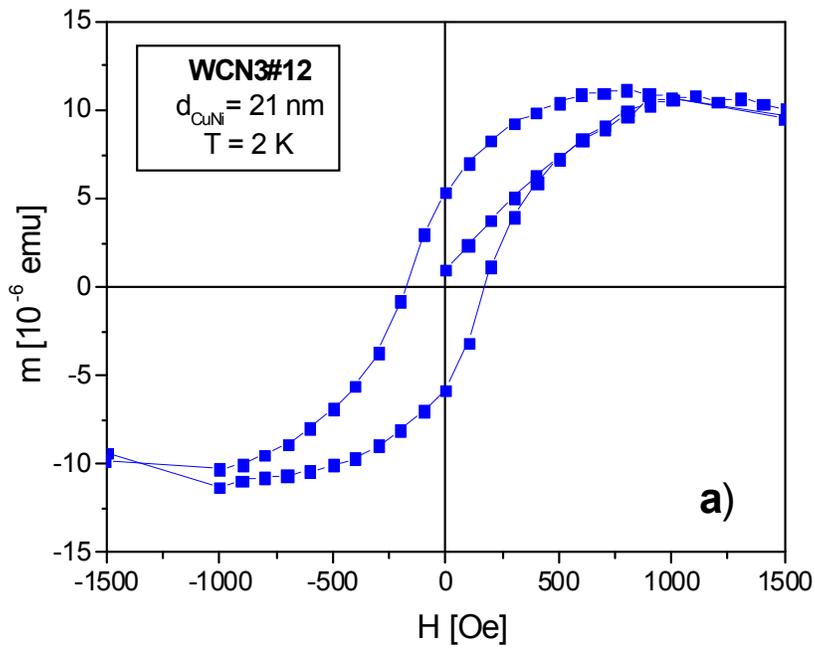

Fig. 4a.

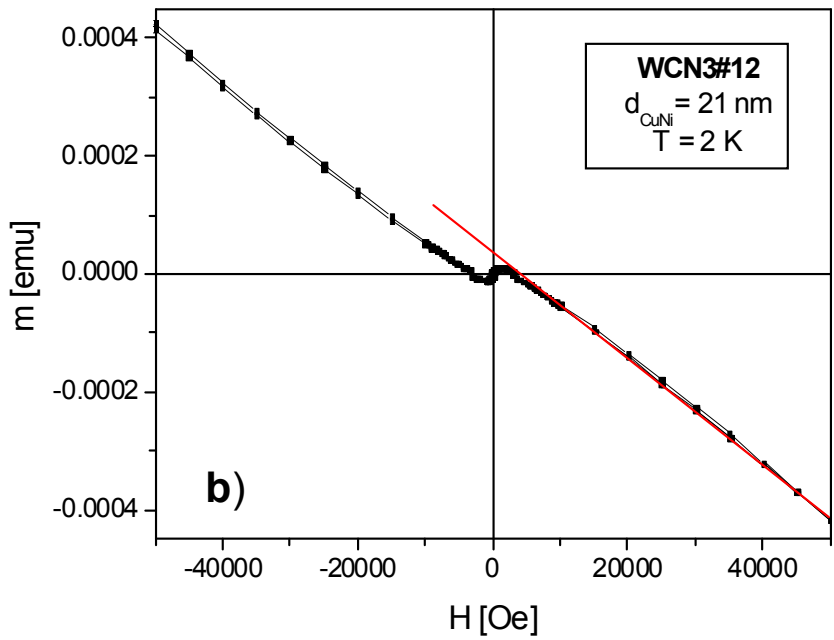

Fig. 4b)

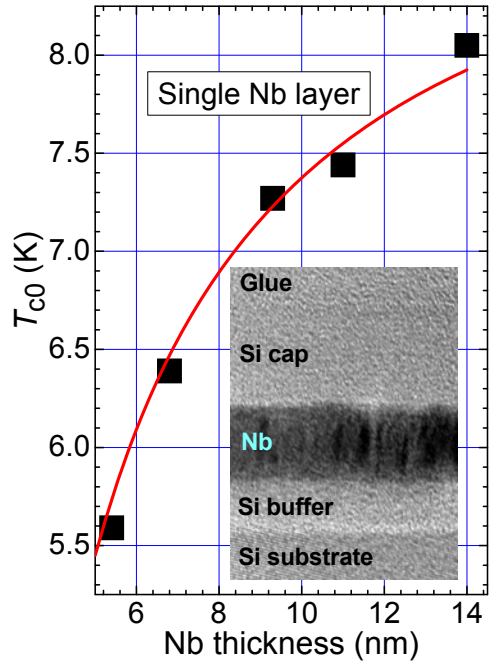

Fig.5

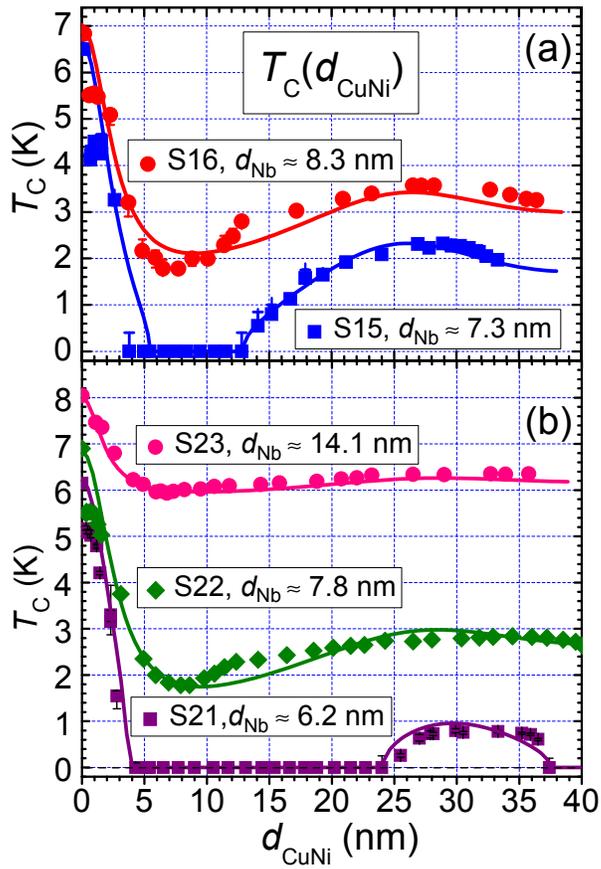

Fig.6

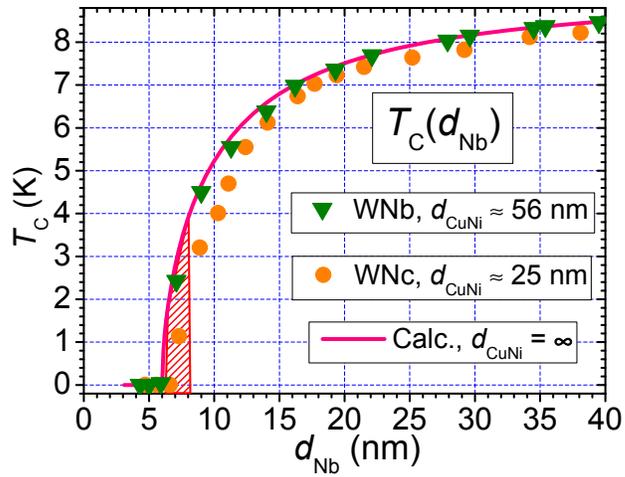

Fig.7

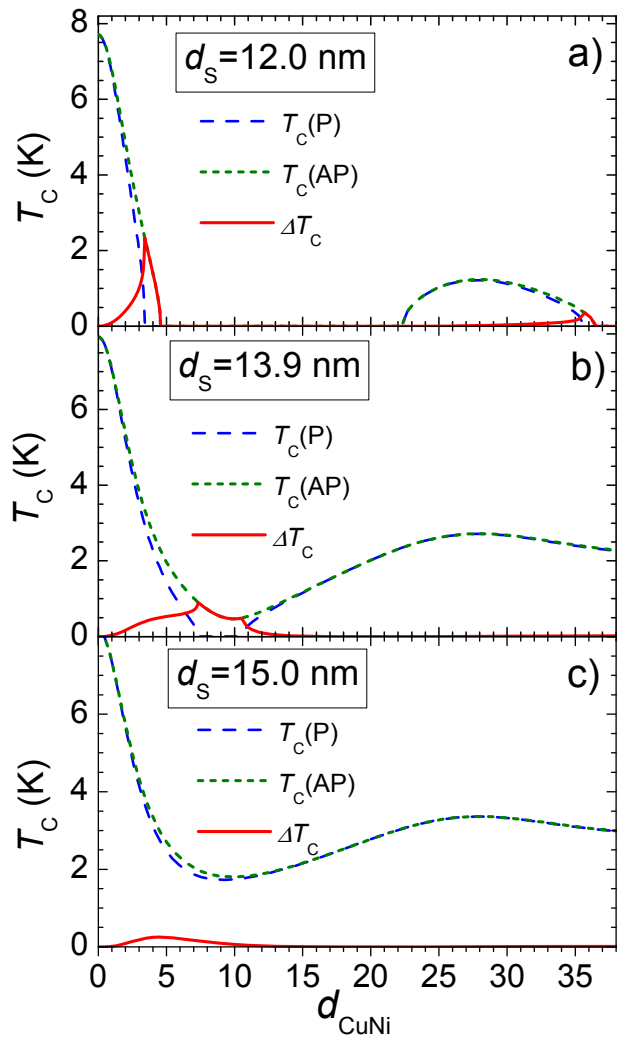

Fig.8

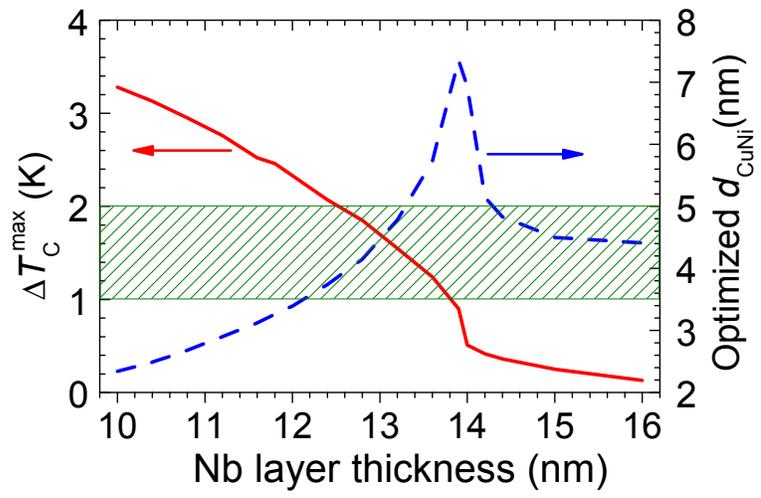

Fig. 9